\documentstyle[12pt]{article}

\advance \topmargin by -\headheight
\advance \topmargin by -\headsep

\evensidemargin \oddsidemargin
\def\ie{{\em i.e.}}

\def\ie{\hbox{\it i.e.}}

\def\CC{{\mathchoice
{\rm C\mkern-8mu\vrule height1.45ex depth-.05ex 
width.05em\mkern9mu\kern-.05em}
{\rm C\mkern-8mu\vrule height1.45ex depth-.05ex 
width.05em\mkern9mu\kern-.05em}
{\rm C\mkern-8mu\vrule height1ex depth-.07ex 
width.035em\mkern9mu\kern-.035em}
{\rm C\mkern-8mu\vrule height.65ex depth-.1ex 
width.025em\mkern8mu\kern-.025em}}}

\def\RR{{\rm I\kern-1.6pt {\rm R}}}

\def\ZZ{{\rm Z}\kern-3.8pt {\rm Z} \kern2pt}

\def\np{Nucl. Phys.}
\def\pl{Phys. Lett.}

\def\pr{Phys. Rev.}

\def\jhep{J. High Energy Phys.}

\newcommand{\beq}{\begin{equation}}
\newcommand{\eeq}{\end{equation}}
\newcommand{\rc}{\nonumber\\}
\newcommand{\bear}{\begin{eqnarray}}
\newcommand{\eear}{\end{eqnarray}}

\newfont{\namefont}{cmr10}
\newfont{\addfont}{cmti7 scaled 1440}
\newfont{\boldmathfont}{cmbx10}
\newfont{\headfontb}{cmbx10 scaled 1728}
\begin{document}
\begin{titlepage}

\begin{center} \Large \bf Flux Stabilization of D-branes in
 a non-threshold bound state background
\end{center}

\vskip 0.3truein
\begin{center} 
J. M. Camino
\footnote{e-mail:camino@fpaxp1.usc.es}
and 
A.V. Ramallo
\footnote{e-mail:alfonso@fpaxp1.usc.es}

\vspace{0.3in}

Departamento de F\'\i sica de
Part\'\i culas, \\ Universidad de Santiago\\
E-15706 Santiago de Compostela, Spain. 
\vspace{0.3in}

\end{center}
\vskip 1truein

\begin{center}
\bf ABSTRACT
\end{center} 

We study some configurations of brane probes which are partially wrapped on
spheres transverse to a stack of non-threshold bound states. The latter are 
represented by the corresponding supergravity background. Two cases are
studied: D(10-p)-branes in the background of (D(p-2), Dp) bound states and
D(8-p)-branes in the (NS5, Dp) geometry. By using suitable flux quantization
rules of the worldvolume gauge field, we determine the stable configurations of
the probe. The analysis of the energy and supersymmetry of these configurations 
reveals that they  can be interpreted as bound states of lower
dimensional objects polarized into a D-brane.

\vskip4.5truecm
\leftline{US-FT-1/02\hfill January 2002}
\leftline{hep-th/0201138}
\smallskip
\end{titlepage}
\setcounter{footnote}{0}


\setcounter{equation}{0}
\section{Introduction}
\medskip
A brane probe wrapped on a sphere in such a way that it captures some flux of a
background gauge field may be stable against shrinking if it is located at a
discrete set of positions determined by a flux quantization rule. This flux
stabilization phenomenon, discovered in refs. \cite{Bachas,Pavel} for the
Neveu-Schwarz (NS) gauge field, was generalized in refs. \cite{PR,flux} for
Ramond-Ramond (RR) gauge field fluxes. In these papers the brane probe is
(partially) wrapped on a sphere $S^d$, which is defined as the set of points of a
$(d+1)$-dimensional sphere which have the same latitude, \ie\ with the same
polar angle. The flux quantization rules and the minimal energy
condition fix this angle, whose value must belong to a finite set. In
particular, in ref. \cite{flux} a generalization of the flux quantization rule
of \cite{Bachas} is proposed, and new sets of angles and energies are obtained. 
These brane configurations admit the interpretation of bound states of strings
polarized by the background fields by means of the Myers mechanism
\cite{Myers}. 

In this paper we shall study the flux stabilization in backgrounds created by
stacks of non-threshold bound states. Two cases will be analyzed: a
D(10-p)-brane probe in the background of  (D(p-2), Dp) bound states and a
D(8-p)-brane moving in the (NS5, Dp) geometry. In these two cases we will
characterize the stable configurations and we will determine their energy. From
these results we will conclude that our configurations can be regarded as bound
states of fundamental strings or (F, D(6-p))-branes in the (D(p-2), Dp) or 
(NS5, Dp) case respectively. We shall confirm this conclusion by determining
the supersymmetry preserved by our solutions. 

\setcounter{equation}{0}
\section{Flux quantization in the (D(p-2), Dp) background}
\medskip
The string frame metric $ds^2$ and the dilaton $\phi$ generated by a stack of 
(D(p-2), Dp) bound states $(p\ge 2)$ are \cite{NCback}:
\bear
ds^2&=&f_p^{-1/2}\,\Big[\,-(\,dx^0\,)^2\,+\,\cdots\,+\,(\,dx^{p-2}\,)^2\,+\,
\,h_p\,\Big((\,dx^{p-1}\,)^2\,+\,(\,dx^{p}\,)^2\Big)\,\Big]\,+\rc\rc
&&+\,f_p^{1/2}\, \Big[\,dr^2\,+\,r^2\,d\Omega_{8-p}^2\,\Big]\,\,,\rc\rc
e^{\tilde\phi}&=&f_p^{{3-p\over 4}}\,\,h_p^{1/2}\,\,,
\label{uno}
\eear
where $\tilde\phi=\phi\,-\,\phi(r\rightarrow\infty)$ and $d\Omega_{8-p}^2$ and
$r$ are, respectively, the line element of a unit $(8-p)$-sphere and   
a radial coordinate which measures the distance to the bound state. 
The functions $f_p$ and $h_p$ in eq. (\ref{uno}), in the near-horizon region of
the metric, are:
\beq
f_p={R^{7-p}\over r^{7-p}}\,\,,
\,\,\,\,\,\,\,\,\,\,\,\,\,\,\,\,\,\,\,\,\,\,\,\,\,\,\,\,\,\,
h_p^{-1}=\sin^2\varphi\,f_p^{-1}\,+\,\cos^2\varphi\,\,,
\label{dos}
\eeq
with $\varphi$ being a constant angle characteristic  of the bound state and,
if $N$ denotes the number of branes of the stack,  $R$ is given by:
\beq
R^{7-p}\,\cos\varphi\,=\,N\,g_s\,2^{5-p}\,\pi^{{5-p\over 2}}\,
(\,\alpha\,'\,)^{{7-p\over 2}}\,\,
\Gamma\Bigl(\,{7-p\over 2}\Bigr)\,\,.
\label{tres}
\eeq
In eq. (\ref{tres}) $g_s$ is the string coupling
constant ($g_s=e^{\phi(r\rightarrow\infty)}$) and $\alpha\,'$ is the Regge
slope. It is clear from the form of the metric  (\ref{uno}) that
the Dp-brane of the background extends along the directions $x^0\cdots x^p$,
whereas the  D(p-2)-brane component lies along $x^0\cdots x^{p-2}$. This
supergravity solution also contains a NSNS two-form potential $B$:
\beq
 B=\tan\varphi\,h_p\,f_p^{-1}\,\,dx^{p-1}\wedge dx^{p}\,\,,
\label{cuatro}
\eeq
and is charged under two RR field strengths $F^{(p)}$ and 
$F^{(p+2)}$, whose components along the directions parallel to the bound state
are:
\beq
F^{(p)}_{x^0,x^1,\cdots, x^{p-2},r}
=\sin\varphi\,\partial_r\,f_p^{-1}\,\,,
\,\,\,\,\,\,\,\,\,\,\,\,\,\,\,\,\,\,\,\,\,\,\,\,\,\,\,\,\,\,
F^{(p+2)}_{x^0,x^1,\cdots, x^{p},r}=\cos\varphi\,
h_p\partial_r\,f_p^{-1}\,.
\label{cinco}
\eeq
From the components of the RR fields displayed in eq. (\ref{cinco}) one can
compute the components of the Hodge dual fields ${}^*\,F^{(p)}$ and 
${}^*\,F^{(p+2)}$ along the directions transverse to the bound state. Clearly 
${}^*\,F^{(p)}$ is a $(10-p)$-form whereas ${}^*\,F^{(p+2)}$ is a 
$(8-p)$-form. Then, they can be represented by means of two RR potentials 
$C^{(9-p)}$ and $C^{(7-p)}$ which are, respectively, a $(9-p)$-form and 
a $(7-p)$-form. In order to write the relevant components of these potentials,
let us parametrize the $S^{8-p}$ transverse sphere by means of the spherical
angles $\theta^1$, $\theta^2$, $\cdots$, $\theta^{8-p}$ and let 
$\theta\equiv\theta^{8-p}$ be the polar angle measured from one of the poles of
the sphere ($0\le\theta\le\pi$). Then, the $S^{8-p}$ line element 
$d\Omega_{8-p}^2$ can be decomposed as:
$d\Omega_{8-p}^2\,=\,d\theta^2\,+\,(\,{\rm sin}\,\theta)^{2}\,\,
d\Omega_{7-p}^2$, where $d\Omega_{7-p}^2$ is the metric of the constant
latitude $(7-p)$-sphere. Let us now define the functions  $C_p(\theta)$  as the
solutions of the initial value problems:
\beq
{d\over d\theta}\, C_p(\theta)\,=\,-(7-p)\,({\rm sin}\,\theta)^{7-p}\,\,,
\,\,\,\,\,\,\,\,\,\,\,\,\,\,\,\,\,\,\,\,\,\,\,\,\,\,\,\,
C_p(0)\,=\,0\,\,,
\label{seis}
\eeq
which can be straightforwardly solved by elementary integration. In terms of the
$C_p(\theta)$'s, the components of the RR potentials in which we are interested
in are:
\bear
C^{(7-p)}_{\theta^1,\cdots,\theta^{7-p}}
&=&-\,\cos\varphi\,R^{7-p}\,C_p(\theta)\,
\sqrt{\hat g^{(7-p)}}\,\,, \rc
C^{(9-p)}_{x^{p-1},x^p,\theta^1,\cdots, \theta^{7-p}}
&=&-\,\sin\varphi\,R^{7-p}\,
h_p\,f_p^{-1}\,C_p(\theta)\,
\sqrt{\hat g^{(7-p)}}\,\,.
\label{siete}
\eear
In eq. (\ref{siete}) $\hat g^{(7-p)}$ is the determinant of the metric of the 
unit $S^{7-p}$ sphere. 

Let us now place a D(10-p)-brane probe in the (D(p-2), Dp) geometry. The action
of such a brane probe is the sum of a Dirac-Born-Infeld and a Wess-Zumino term,
namely \cite{swedes}:
\bear
S\,=\,
-T_{10-p}\,\int d^{\,11-p}\xi\,e^{-\tilde\phi}\,
\sqrt{-{\rm det}\,(\,g\,+\,{\cal F}\,)}\,+\,
T_{10-p}\int\Bigg[\,\,C^{(9-p)}\,\wedge\,{\cal F}+\,
{1\over 2}\, C^{(7-p)}\,\wedge\,
{\cal F}\wedge\,{\cal F}\,\Bigg],\rc
\label{ocho}
\eear
where $g$ is the induced metric on the worldvolume of the brane probe, 
$T_{10-p}$ is the tension of the  D(10-p)-brane and 
${\cal F}=F-B$, with $F$ being a $U(1)$ worldvolume gauge field and $B$ the
NSNS gauge potential (actually its pullback to the probe worldvolume).  We want
to find stable configurations in which the probe is partially wrapped on the 
$S^{7-p}$ constant latitude sphere. From the analysis performed in ref.
\cite{flux} for RR backgrounds, it follows that we must extend the probe along
the radial coordinate and switch on an electric worldvolume field along this
direction. Moreover, our background has a $B$ field with non-zero components
along the 
$x^{p-1}x^{p}$ plane. Then, in order to capture the flux of the $B$ field, we
must also extend our D(10-p)-brane probe along the $x^{p-1}x^{p}$ directions.
Therefore, the natural set of worldvolume coordinates 
$\xi^{\alpha}$ ($\alpha=0,\cdots,10-p$) is
$\xi^{\alpha}\,=\,(t,x^{p-1},x^{p},r,\theta^1,\cdots,\theta^{7-p})$,  where
$t\equiv x^{0}$.
Moreover, we will adopt the following ansatz for the field ${\cal F}$:
\beq
{\cal F}\,=\,F_{0,r}\,dt\wedge dr\,
+\,{\cal  F}_{p-1,p}\,dx^{p-1}\wedge dx^{p}\,\,.
\label{nueve}
\eeq
Notice that ${\cal  F}_{p-1,p}$ gets a contribution from the pullback of $B$,
namely:
\beq
{\cal  F}_{p-1,p}\,=\,  F_{p-1,p}\,-\,h_pf_p^{-1}\tan\varphi\,\,.
\label{diez}
\eeq
It is interesting to point out that the components of ${\cal  F}$ in eq. 
(\ref{nueve}) are precisely those which couple to the RR potentials 
(\ref{siete}) in the Wess-Zumino term of the action. With the election of
worldvolume coordinates we have made above, the embedding of the brane probe in
the transverse space is encoded in the dependence of the polar angle $\theta$ on
the  $\xi^{\alpha}$'s. Although we are interested in configurations in which
$\theta$ is constant, we will consider first the more general situation in
which $\theta=\theta(r)$. It is easy to compute the lagrangian in this case.
One gets:
\beq
L\,=\,\int dx^{p-1}dx^p\int_{S^{7-p}}\,d^{7-p}\theta\,
\sqrt{\hat g}\,\int\,drdt\,{\cal L}(\theta, F)\,\,,
\label{once}
\eeq
where $\hat g\equiv \hat g^{(7-p)}$ and the lagrangian density 
${\cal L}(\theta, F)$ is:
\bear
{\cal L}(\theta, F)\,=\,-T_{10-p}\,R^{7-p}&&\Big[\,
(\sin\theta)^{7-p}\,
\sqrt{h_pf_p^{-1}+h_p^{-1}{\cal  F}_{p-1,p}^2}\,\,
\sqrt{1+r^2\theta'^2-F_{0,r}^2}\,+\,\rc\rc
&&+\,\cos\varphi F_{p-1,p}\,F_{0,r}\,C_p(\theta)\,
\Big]\,\,.
\label{doce}
\eear
We want to find solutions of the equations of motion derived from 
${\cal L}(\theta, F)$ in which both the angle $\theta$ and the worldvolume
gauge field are constant. The equation of motion for
$\theta$ with $\theta=\bar\theta={\rm constant}$ reduces to:
\beq
\cos\bar\theta\,\sqrt{1-F_{0,r}^2}\,
\sqrt{h_pf_p^{-1}+h_p^{-1}{\cal  F}_{p-1,p}^2}\,-\,
\sin\bar\theta\,\cos\varphi\,F_{0,r}\,F_{p-1,p}\,=\,0\,\,.
\label{trece}
\eeq
If $F_{0,r}$ and $F_{p-1,p}$ are constant, eq. (\ref{trece}) is only consistent
when its left-hand side is independent of $r$. However, the square root
involving ${\cal  F}_{p-1,p}$ does depend on $r$ in general. Actually, after a
simple calculation one can verify that:
\beq
h_pf_p^{-1}+h_p^{-1}{\cal F}_{p-1,p}^2\,=
\,\cos^2\varphi\,F_{p-1,p}^2\,+
\,f_p^{-1}\,\Big(\,F_{p-1,p}\sin\varphi\,-\,
{1\over \cos\varphi}\Big)^{2}\,\,.
\label{catorce}
\eeq
By inspecting the right-hand side of eq. (\ref{catorce}) one immediately
concludes that it is only independent of $r$ when $F_{p-1,p}$ takes the value:
\beq
F_{p-1,p}\,=\,{1\over \sin\varphi\cos\varphi}
\,=\,2\csc (2\varphi)\,\,.
\label{quince}
\eeq
Plugging back this value of $F_{p-1,p}$ into eq. (\ref{trece}), one gets that 
$F_{0,r}\,=\,\cos\bar\theta$. In order to determine the allowed values of 
$\bar\theta$, and therefore of $F_{0,r}$, we need to impose a quantization
condition. Let us consider again a configuration in which $\theta=\theta(r)$
and assume that $F_{p-1,p}$ is given by eq. (\ref{quince}). Moreover, let us
introduce a quantization volume ${\cal V}$ in the $x^{p-1}x^{p}$ plane which
corresponds to one unit of flux, namely:
\beq
\int_{{\cal V}}\,dx^{p-1}\,dx^{p}\,F_{p-1,p}\,=\,
{2\pi\over T_f}\,\,,
\label{dseis}
\eeq
where $T_{f}\,=\,( 2\pi\alpha\,'\,)^{-1}\,$. 
By using the constant value of $F_{p-1,p}$ written in eq. (\ref{quince}), one
gets that ${\cal V}$ is given by:
\beq
{\cal V}\,=\,2\pi^2\alpha'\sin(2\varphi)\,.
\label{dsiete}
\eeq
We now determine $F_{0,r}$ by imposing the quantization condition of ref. 
\cite{flux} on the volume ${\cal V}$, \ie:
\beq
\int_{{\cal V}}\,dx^{p-1}\,dx^{p}\,
\int_{S^{7-p}}\,\,d^{7-p}\theta\,\sqrt{\hat g}\,
{\partial \,{\cal L}\over\partial F_{0,r}}\,=\, n\,T_{f}\,\,,
\label{docho}
\eeq
with $n\in\ZZ$. By using the explicit form of ${\cal L}$ and $F_{p-1,p}$, one
can easily compute the left-hand side of the quantization condition 
(\ref{docho}):
\bear
\int_{{\cal V}}\,dx^{p-1}\,dx^{p}\,
\int_{S^{7-p}}\,\,d^{7-p}\theta\,\sqrt{\hat g}\,
{\partial \,{\cal L}\over\partial F_{0,r}}\,=\, 
{T_{10-p}\,\Omega_{7-p}\,R^{7-p}\,{\cal V}\over \sin\varphi}
\Biggl[\,
{F_{0,r}\,\,({\rm sin}\,\theta)^{7-p}\,\over 
\sqrt{1\,+\,r^2\,\theta\,'^{\,2}\, -\,F_{0,r}^2}}\,\,-\,\,
C_p(\theta)\,\Biggr]\,\,,\rc
\label{dnueve}
\eear
where $\Omega_{7-p}$ is the volume of the unit $(7-p)$-sphere and we have
assumed that $F_{0,r}$ does not depend on $\theta^1,\cdots\theta^{7-p}$. It is
not difficult now to find $F_{0,r}$ as a function of $\theta(r)$ and the
quantization integer $n$. First of all, let us notice that the global 
coefficient of the right-hand side of (\ref{dnueve}) is:
\beq
{T_{10-p}\,\Omega_{7-p}\,R^{7-p}\,{\cal V}\over \sin\varphi}
\,=\,
{NT_f\over 2\sqrt{\pi}}\,
{\Gamma\Bigl(\,{7-p\over 2}\Bigr)\over
\Gamma\Bigl(\,{8-p\over 2}\Bigr)}\,\,. 
\label{veinte}
\eeq
Secondly, let us  define a new function ${\cal C}_{p,n}(\theta)$ as:
\beq
{\cal C}_{p,n}(\theta)\,=\,C_p(\theta)\,+\,2\,\sqrt{\pi}\,\,
{\Gamma\Bigl(\,{8-p\over 2}\Bigr)\over
\Gamma\Bigl(\,{7-p\over 2}\Bigr)}\,\,
{n\over N}\,\,.
\label{vuno}
\eeq
Then, the electric field is given by:
\beq
F_{0,r}\,=\,\sqrt{\,
{1\,+\,r^2\,\theta\,'^{\,2}\over 
{\cal C}_{p,n}(\theta)^2\,+\,
({\rm sin}\,\theta)^{2(7-p)}}}\,\,{\cal C}_{p,n}(\theta)\,\,.
\label{vdos}
\eeq
Once $F_{0,r}$ is known, we can obtain the hamiltonian $H$ by means of a
Legendre transformation:
\beq
H\,=\,\int\,dx^{p-1}\,dx^{p}\,
\int_{S^{7-p}}\,\,d^{7-p}\theta\,\,\sqrt{\hat g}\,
\int dr\,\Big[\, F_{0,r}\,{\partial \,{\cal L}\over\partial
F_{0,r}}\,-\, {\cal L}\,\Big]\,\,.
\label{vtres}
\eeq
By using eqs. (\ref{doce}) and (\ref{vdos}), one easily obtains the following
expression of $H$:
\beq
H\,=\,{T_{10-p}\,\Omega_{7-p}\,R^{7-p}\over \sin\varphi}\,
\int\,dx^{p-1}\,dx^{p}\,\int dr\,
\sqrt{1\,+\,r^2\,\theta\,'^{\,2}}\,\,
\sqrt{\,({\rm sin}\,\theta)^{2(7-p)}+\,
\Big({\cal C}_{p,n}(\theta)\Big)^2}\,\,. 
\label{vcuatro}
\eeq
The constant $\theta$ solutions of the equation of motion are those which
minimize $H$ for $\theta\,'=0$. In order to characterize these solutions, let
us define the functions:
\beq
\Lambda_{p,n}(\theta)\,\equiv\,
({\rm sin}\,\theta)^{6-p}\,{\rm cos}\,\theta\,-\,
{\cal C}_{p,n}(\theta)\,\,.
\label{vcinco}
\eeq
Then, the vanishing of $\partial H/\partial \theta$ for $\theta\,'=0$ occurs
when $\theta=\bar\theta_{p,n}$, where $\bar\theta_{p,n}$ is determined by the
condition:
\beq
\Lambda_{p,n}(\bar\theta_{p,n})\,=\,0\,\,.
\label{vseis}
\eeq
The properties of the functions $\Lambda_{p,n}(\bar\theta_{p,n})$ and the
solutions of eq. (\ref{vseis}) have been studied in ref. \cite{flux}, where
it was proved that there exists a unique solution  $\bar\theta_{p,n}$ in the
interval $[0,\pi]$ for $p\le 5$ and $0\le n\le N$. The values of the angles 
$\bar\theta_{p,n}$ for $p=4,5$ can be given in analytic form, namely
$\bar\theta_{5,n}=n\pi/N$ and $\bar\theta_{4,n}=\arccos[1-2n/N]$. Moreover, for
all values of $p\le 5$, $\bar\theta_{p,0}=0$ and $\bar\theta_{p,N}=\pi$, which
correspond to singular configurations in which the brane probe collapses at one
of the poles of the $S^{7-p}$ sphere. Excluding these points, there are exactly
$N-1$ angles  which minimize the energy. The corresponding electric field is
$F_{0,r}=\cos\bar\theta_{p,n}$. Furthermore, if we integrate $x^{p-1}$
and $x^{p}$ in eq. (\ref{vcuatro}) over the quantization volume ${\cal V}$, we
obtain  the energy $H_{p,n}$ of these solutions on the volume ${\cal V}$, which 
can be written as:
\beq
H_{p,n}\,=\,\int\,dr\,{\cal E}_{p,n}\,\,,
\label{vsiete}
\eeq
where the constant energy density ${\cal E}_{p,n}$ is given by:
\beq
{\cal E}_{p,n}\,=\,
{NT_f\over 2\sqrt{\pi}}\,
{\Gamma\Bigl(\,{7-p\over 2}\Bigr)\over
\Gamma\Bigl(\,{8-p\over 2}\Bigr)}\,\,
({\rm sin}\,\bar\theta_{p,n})^{6-p}\,\,.
\label{vocho}
\eeq
The expression of ${\cal E}_{p,n}$  in (\ref{vocho}) is the same as that found
in ref. \cite{flux} for a Dp-brane background. It was argued in \cite{flux}
that ${\cal E}_{p,n}$ can be interpreted as the energy density of a bound state
of $n$ fundamental strings. Actually,  one can verify from (\ref{vocho}) that 
${\cal E}_{p,n}\le n T_f$ and that ${\cal E}_{p,n}\rightarrow n T_f$ in the
semiclassical limit $N\rightarrow\infty$. Thus, we are led to propose that the
states we have found are, in fact, a bound state of polarized fundamental
strings stretched along the radial direction and distributed over the
$x^{p-1}x^p$ plane in such a way that there are $n$ fundamental strings in the
volume ${\cal V}$. Notice that ${\cal V}\rightarrow 0$ when 
$\varphi\rightarrow 0$ and, thus, the bound state becomes point-like in the 
$x^{p-1}x^p$ directions as $\varphi\rightarrow 0$. This fact is in agreement
with ref. \cite{flux}, since, in this limit, the (D(p-2), Dp) background
becomes the Dp-brane geometry. 

In order to confirm the interpretation of our results given above, let us study
the supersymmetry preserved by our brane probe configurations. In general, 
the number of supersymmetries preserved by a D-brane is the number of
independent solutions of the equation 
$\Gamma_{\kappa}\,\epsilon\,=\,\epsilon$,
where $\epsilon$ is a Killing spinor of the background and $\Gamma_{\kappa}$
is the so-called $\kappa$-symmetry matrix \cite{bbs}. For simplicity, we shall
restrict ourselves to the analysis of the $p=3$ case, \ie\ for the (D1, D3)
background. The Killing spinors in this case have the form:
\beq
\epsilon\,=\,e^{{\alpha\over 2}\,\Gamma_{\underline{x^{2}x^{3}}}\,
\sigma_3}\,\,\tilde\epsilon\,\,,
\label{vnueve}
\eeq
where $\Gamma_{\underline {x^{M_1}x^{M_2}}\cdots}$ are antisymmetrized products
of ten-dimensional constant gamma matrices, $\tilde\epsilon$ is a spinor which
satisfies 
$(i\sigma_2)\,
\Gamma_{\underline{x^0\cdots x^3}}\,\,\tilde\epsilon\,\,=\,\,
\tilde\epsilon$ and $\alpha$ is given by: 
\beq
\sin\alpha\,=\,f_3^{-{1\over 2}}\,h_3^{{1\over 2}}\sin\varphi\,\,,
\,\,\,\,\,\,\,\,\,\,\,\,\,\,\,\,\,\,\,\,\,\,
\cos\alpha\,=\,h_3^{{1\over 2}}\cos\varphi\,\,.
\label{treinta}
\eeq
Moreover, the $\kappa$-symmetry matrix \cite{bbs} of the D7-brane probe can be
put as:
\beq
\Gamma_{\kappa}\,=\,{i\sigma_2\over 
\sqrt{1\,-\,F_{0,r}^2}}\,
\Big[\,F_{0,r}\,-\,\Gamma_{\underline {x^0r}}\sigma_3\,\Big]\,
e^{-\eta\Gamma_{\underline {x^2x^3}}\sigma_3}\,\Gamma_{*}\,\,,
\label{tuno}
\eeq
where $\Gamma_{*}\,=\,\Gamma_{\underline {\theta^1\cdots\theta^4}}$ and $\eta$
is:
\beq
\sin\eta\,=\,{f_3^{-{1\over 2}}\,h_3^{{1\over 2}}\over 
\sqrt{h_3f_3^{-1}+h_3^{-1}{\cal F}_{2,3 }^2}
}\,\,,
\,\,\,\,\,\,\,\,\,\,\,\,\,\,\,\,\,\,\,\,\,\,
\cos\eta\,=\,{{\cal F}_{2,3 }\,h_3^{-{1\over 2}}\over 
\sqrt{h_3f_3^{-1}+h_3^{-1}{\cal F}_{2,3 }^2}
}\,\,.
\label{tdos}
\eeq
For our configurations, in which $F_{2,3 }$ is given by eq. (\ref{quince}), the
angles $\alpha$  and $\eta$ of eqs. (\ref{treinta}) and (\ref{tdos})
are equal,  and the
$\Gamma_{\kappa}\,\epsilon\,=\,\epsilon$ condition becomes:
\beq
{1\over \sin\theta}\,\Big[\,\cos\theta\,+\,
\Gamma_{\underline {x^0r}}\,\sigma_3\,\Big]\,
\Gamma_{\underline {\theta r}}\,\tilde\epsilon\,=\,
\tilde\epsilon\,\,.
\label{ttres}
\eeq
Notice that, in order to derive (\ref{ttres}) we have used that
$F_{0,r}=\cos\theta$ in eq. (\ref{tuno}). Moreover, 
introducing the $\theta$-dependence of the spinors, \ie\ 
$\tilde\epsilon\,=\,\exp[-{\theta\over 2}\,\Gamma_{\underline {\theta r}}\,]\,
\hat\epsilon$, with $\hat\epsilon$ independent of $\theta$, 
we get the following condition on $\hat\epsilon$:
\beq
\Gamma_{\underline {x^0r}}\,\sigma_3\hat\epsilon\,=\,\hat\epsilon\,\,,
\label{tcuatro}
\eeq
which can be rewritten as:
\beq
\Gamma_{\underline {x^0r}}\,\sigma_3
\epsilon_{{\,\big |}_{\theta=0}}=\,\epsilon_{{\,\big |}_{\theta=0}}\,\,,
\label{tcinco}
\eeq
which certainly corresponds to a system of fundamental strings in the radial
direction. Notice that the point $\theta=0$ can be regarded as the ``center of
mass" of the expanded fundamental strings.

\setcounter{equation}{0}
\section{Flux quantization in the (NS5, Dp) background}
\medskip
We will now consider a background \cite{NS5Dp} generated by a stack of $N$ bound
states of NS5-branes and Dp-branes for $1\le p\le 5$. The bound state is
characterized by two coprime integers $l$ and $m$ which, respectively, 
determine the number of NS5-branes and Dp-branes which form the bound state. We
shall combine $l$ and
$m$ to form the quantity  $\mu_{(l,m)}\,=\,l^2\,+\,m^2\,g_s^2$. Moreover, for a
stack of $N$ (NS5, Dp) bound states we define 
$R^2_{(l,m)}\,=\,N\,\Bigl[\,\mu_{(l,m)}\,\Bigr]^{{1\over 2}}
\,\alpha'$, in terms of which the near-horizon  harmonic function $H_{(l,m)}(r)$
is defined as:
\beq
H_{(l,m)}(r)\,=\,{R^2_{(l,m)}\over r^2}\,\,.
\label{tseis}
\eeq
The metric of this background in the string frame is:
\bear
&&ds^2=\Big[\,h_{(l,m)}(r)\,\Big]^{-{1\over 2}}\,\,
\Bigg[\,\,
\Bigl[\,H_{(l,m)}(r)\,\Bigr]^{-{1\over 4}}\,\,
\Big(\,-dt^2\,+\,(dx^{1})^2\,+\,\cdots\,+\,(dx^{p})^2\,\Big)\,+\,\rc\rc
&&+\,h_{(l,m)}(r)\Bigl[\,H_{(l,m)}(r)\,\Bigr]^{{1\over 4}}\,\,
\Big(\,(dx^{p+1})^2\,+\cdots\,+\,(dx^{5})^2\,
\Big)\,+\,
\Bigl[\,H_{(l,m)}(r)\,\Bigr]^{{3\over 4}}\,\,
(\,dr^2\,+\,r^2\,d\Omega_{3}^2\,)\,\Bigg]\,,\rc
\label{tsiete}
\eear
where the function $h_{(l,m)}(r)$ is given by:
\beq
h_{(l,m)}(r)\,=\,{\mu_{(l,m)}\over
l^2\,\,\Bigl[\,H_{(l,m)}(r)\,\Bigr]^{{1\over 2}}
\,+\,m^2\,g_s^2\,\,
\Bigl[\,H_{(l,m)}(r)\,\Bigr]^{-{1\over 2}}}\,\,.
\label{tocho}
\eeq

The NS5-branes of this background extend along the $tx^1\cdots x^5$
coordinates, whereas the Dp-branes lie along $tx^1\cdots x^p$ and are smeared
in the $x^{p+1}\cdots x^5$ coordinates. The integers $l$ and $m$ represent,
respectively, the number of NS5-branes in the bound state and the number of 
Dp-branes in a $(5-p)$-dimensional volume 
${\cal V}_p\,=\,(2\pi \sqrt{\alpha'})^{5-p}$ in the  $x^{p+1}\cdots x^5$
directions. We shall choose, as in section 2, spherical coordinates, and
we will represent the $S^3$ line element as 
$d\Omega_{3}^2\,=\,d\theta^2\,+\,(\,{\rm sin}\,\theta)^{2}\,\,
d\Omega_{2}^2$. Other fields in this supergravity solution include the 
dilaton:
\beq
e^{-\phi}\,=\,g_s^{-1}\,\Bigl[\,H_{(l,m)}(r)\,\Bigr]^{{p-5\over 8}}\,
\Bigl[\,h_{(l,m)}(r)\,\Bigr]^{{p-1\over 4}}\,\,,
\label{tnueve}
\eeq
the NSNS potential $B$:
\beq
B\,=\,-l\,N\alpha'\,C_5(\theta)\,\epsilon_{(2)}\,\,,
\label{cuarenta}
\eeq
and two RR potentials $C^{(7-p)}$ and $C^{(5-p)}$, whose relevant components
are:
\bear 
&&C^{(7-p)}_{x^{p+1},\cdots, x^{5},\theta^{1},\theta^{2}}\,=\,
-m\,N\alpha'\,C_5(\theta)\,\sqrt{\hat g^{(2)}}\,\,,\rc\rc
&&C^{(5-p)}_{x^{p+1},\cdots ,x^{5}}\,=\,
{lm\over \mu_{(l,m)}}\,
\Big(\,\Bigl[\,H_{(l,m)}(r)\,\bigr]^{{1\over 2}}\,-\,
\bigl[\,H_{(l,m)}(r)\,\bigr]^{-{1\over 2}}
\,\Big)\,h_{(l,m)}(r)\,\,.
\label{cuno}
\eear
In eqs. (\ref{cuarenta}) and (\ref{cuno})
$C_5(\theta)$ is the function defined in eq. (\ref{seis}) for $p=5$, namely 
$C_5(\theta)=\sin \theta\cos\theta\,-\,\theta$,  $\epsilon_{(2)}$ is the volume
element of the constant latitude sphere $S^2$ and $\hat g^{(2)}$ the
determinant of its metric. Moreover, to simplify the equations that follow  we
shall take from now on $g_s=1$ (the dependence on $g_s$ can be easily
restored).

By inspecting the form of the NSNS and RR potentials is eqs. (\ref{cuarenta})
and (\ref{cuno}) one easily realizes that, in order to get flux-stabilized
configurations, one must consider a D(8-p)-brane probe wrapping the $S^2$ and
extended along $r,x^{p+1},\cdots,x^{5}$. The action of such a probe is:
\bear
S\,&=&\,
-T_{8-p}\,\int d^{\,9-p}\xi\,e^{-\tilde\phi}\,
\sqrt{-{\rm det}\,(\,g\,+\,{\cal F}\,)}\,+\,\rc\rc
&&+\,T_{8-p}\,\int\Big[\,C^{(7-p)}\wedge {\cal F}\,+\,
{1\over 2}\,C^{(5-p)}\wedge {\cal F} \wedge {\cal F}\,\Big]\,\,.
\label{cdos}
\eear
We shall take in (\ref{cdos}) the following  set of worldvolume 
coordinates 
$ \xi^{\alpha}=(t,x^{p+1},\cdots,x^{5},r,\theta^1,\theta^{2})$ and 
we will consider configurations of the brane probe in which $\theta$ is a
function of $r$.
To determine the worldvolume gauge field $F$,  we first impose 
the flux quantization condition:
\beq
\int_{S^2}\,F\,=\,{2\pi n_1\over T_f}\,\,,
\,\,\,\,\,\,\,\,\,\,\,\,\,\,\,\,\,\,
n_1\in\ZZ\,\,.
\label{ctres}
\eeq
Eq. (\ref{ctres}) can be easily solved, and its solution fixes the magnetic
components of $F$. Actually, if we assume that the electric worldvolume field
has only components along the radial direction, one can write the solution of 
(\ref{ctres}) as 
$F\,=\,\pi n_1\alpha'\epsilon_{(2)}\,+\,F_{0,r}dt\wedge dr$,
which is equivalent to the following expression of ${\cal F}$:
\beq
{\cal F}\,=\,f_{12}(\theta)\epsilon_{(2)}\,+\,F_{0,r}dt\wedge dr\,\,,
\label{ccuatro}
\eeq
with $f_{12}(\theta)$ being:
\beq
f_{12}(\theta)\,\equiv\,lN\alpha'C_5(\theta)\,+\,\pi n_1\alpha'\,\,.
\label{ccinco}
\eeq
Using eq. (\ref{ccuatro}) in (\ref{cdos}) one finds that the 
lagrangian of the system is :
\beq
L\,=\,\int dx^{p+1}\cdots dx^5\int_{S^2}\,d^2\theta\,
\sqrt{\hat g}\,\int\,drdt\,{\cal L}(\theta, F)\,\,,
\label{cseis}
\eeq
where $\hat g\equiv \hat g^{(2)}$ and the lagrangian density is given by:
\bear
{\cal L}(\theta, F)\,&=&\,-T_{8-p}\,\,\Bigg[\,
\sqrt{r^4\,\Bigl[\,H_{(l,m)}(r)\,\Bigr]^{{3\over 2}}\,(\sin\theta)^4
\,+\,h_{(l,m)}(r)\,f_{12}(\theta)^2}\,\times\rc\rc
&&\times\,\sqrt{\Bigl[\,H_{(l,m)}(r)\,\Bigr]^{{1\over 2}}
(1+r^2\theta'^2)\,-\,h_{(l,m)}(r)\,F_{0,r}^2}\,+\,\rc\rc
&&+\,(mN\alpha'C_5(\theta)\,-\,C^{(5-p)} f_{12}(\theta))\,F_{0,r}
\,\Bigg]\,\,.
\label{csiete}
\eear
In eq.  (\ref{csiete}) we have suppressed the indices of the RR potential 
$C^{(5-p)}$. We now impose the following quantization condition \cite{flux}
(see eq. (\ref{docho})):
\beq
\int_{{\cal V}_p}\,\,
dx^{p+1}\cdots dx^5\int_{S^2}\,d^2\theta\,
\sqrt{\hat g}\,\,{\partial {\cal L}\over 
\partial F_{0,r}}\,=\,n_2\,T_f\,\,,
\label{cocho}
\eeq
where $n_2\in\ZZ$ and ${\cal V}_p$ is the quantization volume defined after
eq. (\ref{tocho}) (\ie\ ${\cal V}_p=(2\pi \sqrt{\alpha'})^{5-p}$). 
Eq. (\ref{cocho}) allows to obtain
$F_{0,r}$ as a function of $\theta(r)$ and of the quantization integers $n_1$
and $n_2$.  By means of a Legendre transformation one can get the form of
the hamiltonian of the system. After some calculation one arrives at:
\bear
&&H\,=\,T_{8-p}\Omega_2\,\int dx^{p+1}\cdots dx^5
\int dr\sqrt{1\,+\,r^2\theta'^2}\,\times\,\label{cincuenta}\\\rc
&&\times\,\sqrt{R^4_{(l,m)}\,(\sin\theta)^4\,+\,
[\mu_{(l,m)}]^{-1}\big[\,\big(lf_{12}(\theta)+m\Pi(\theta)\big)^2+
H_{(l,m)}(r)\big(mf_{12}(\theta)-l\Pi(\theta)\big)^2\,\big]}\,\,,
\nonumber
\eear
where $\Pi(\theta)$ is the function:
\beq
\Pi(\theta)\,\equiv\,mN\alpha'C_5(\theta)\,+\,
\pi n_2\alpha'\,\,.
\label{ciuno}
\eeq
By inspecting the right-hand side of eq. (\ref{cincuenta}) one immediately
reaches the conclusion that  there exist configurations with  constant $\theta$
which minimize the energy only when $mf_{12}(\theta)\,=\,l\Pi(\theta)$. By
looking at eqs. (\ref{ccinco}) and (\ref{ciuno}) it is immediate to verify that
this condition is equivalent to $mn_1\,=\,ln_2$. Since $l$ and $m$
are coprime, one must have  $n_1\,=\,l\,n\,$, $n_2\,=\,mn\,$ with $n\in\ZZ$. 
Then, our two quantization integers  $n_1$ and $n_2$ are not independent and 
$f_{12}(\theta)$ and $\Pi(\theta)$ are given in terms of $n$ by:
\beq
f_{12}(\theta)\,=\,lN\alpha'\,{\cal C}_{5,n}(\theta)\,,
\,\,\,\,\,\,\,\,\,\,\,\,\,\,\,\,\,\,\,
\Pi(\theta)\,=\,mN\alpha'\,{\cal C}_{5,n}(\theta)\,\,,
\label{cidos}
\eeq
where ${\cal C}_{5,n}(\theta)$ is the function defined in eq. (\ref{vuno}) for
$p=5$. By using eq. (\ref{cidos}) in eq. (\ref{cincuenta}), one gets the
following expression of $H$: 
\bear
H\,=\,T_{8-p}\,\Omega_2\,R_{(l,m)}^2\,
\int dx^{p+1}\cdots dx^5\int\,\, dr
\,\times\,\rc\rc\,
\times\,
\sqrt{1\,+\,r^2\theta'^2}\,\,
\sqrt{(\sin\theta)^4\,+\,
\Big(\,{\cal C}_{5,n}(\theta)\Big)^2}\,\,.
\label{citres}
\eear
By comparing the right-hand side of eq. (\ref{citres}) with that of eq. 
(\ref{vcuatro}) one immediately realizes that the constant angles which 
minimize the energy are the solutions of eq. (\ref{vseis}) for $p=5$, \ie\
$\theta\,=\,\bar\theta_{5,n}\,=\,{n\over N}\,\pi$ with $0\le n\le N$. 
The electric field $F_{0,r}$ which we must have in
the  worldvolume in order to wrap the D(8-p)-brane at
$\theta\,=\,\bar\theta_{5,n}$ is easily obtained from eq. (\ref{cocho}). After
a short calculation one gets that, for a general value of $g_s$, $F_{0,r}$ is
given by:
\beq
\bar F_{0,r}\,=\,{mg_s\over 
\sqrt{l^2+m^2g_s^2}}\,\,
\cos\,\Big[\,{n\over N}\,\pi\Big]\,\,.
\label{cicuatro}
\eeq
Let 
$H_n^{(l,m)}$ be the energy of our configurations and 
${\cal E}_n^{(l,m)}$ the corresponding energy density, whose integral over
$x^{p+1}\cdots x^5 r$ gives $H_n^{(l,m)}$. After a short calculation one easily
proves that:
\beq
{\cal E}_n^{(l,m)}\,=\,
{NT_{6-p}(m,l)\over \pi}\,
\sin\Big[\,{n\over N}\,\pi\Big]\,\,,
\label{cicinco}
\eeq
where, for an arbitrary value of $g_s$, $T_{6-p}(m,l)$ is given by:
\beq
T_{6-p}(m,l)\,=\,
{1\over (2\pi)^{6-p}\,(\alpha')^{{7-p\over 2}} g_s}\,\,
\sqrt{l^2\,+\,m^2\,g_s^2}\,\,.
\label{ciseis}
\eeq
$T_{6-p}(m,l)$  is the tension of a bound state of fundamental strings and
D(6-p)-branes \cite{FDp}. In such a (F, D(6-p))-brane state, $l$ is the number of
D(6-p)-branes, whereas $m$ parametrizes the number of fundamental strings.
Indeed, one can check that $T_{6-p}(0,l)=lT_{6-p}$ and, on the other hand, 
$T_{6-p}(m,0){\cal V}_p=mT_f$, which means that there are $m$ fundamental
strings in the $(5-p)$-dimensional volume ${\cal V}_p$. These strings are
stretched in the radial direction and  smeared in the 
$x^{p+1}\cdots x^5$ coordinates. This interpretation of $T_{6-p}(m,l)$ 
suggests that our configurations with $\theta\,=\,\bar\theta_{5,n}$ are bound
states of (F, D(6-p))-branes. Indeed, since 
${\cal E}_n^{(l,m)}\rightarrow n T_{6-p}(m,l)$ as $N\rightarrow\infty$, the
number of (F, D(6-p))-branes which form our bound state is precisely the
quantization integer $n$. Moreover, we can determine the supersymmetry
preserved by our configuration. This analysis is similar to the one carried out
at the end of section 2. Let us present the result of this study for the 
(NS5, D3) background, which corresponds to taking $p=3$ in our general
expressions. If $\epsilon$ denotes a Killing spinor of the background, only 
those $\epsilon$ which satisfy:
\beq
\Big[\,\cos\alpha\,\Gamma_{\underline{x^0r}}\sigma_3\,+\,
\sin\alpha\,\Gamma_{\underline{x^0x^4x^5r}}(i\sigma_2)\,\Big]\,
\epsilon_{{\,\big |}_{\theta=0}}\,=\,\epsilon_{{\,\big |}_{\theta=0}}\,,
\label{cisiete}
\eeq
generate a supersymmetry transformation which leaves our  configuration
invariant. In eq. (\ref{cisiete}) $\alpha$ is given by:
\beq
\sin\alpha\,=\,{l\over [\mu_{(l,m)}]^{1/2}}\,
H_{(l,m)}^{1/4}\,h_{(l,m)}^{1/2}\,\,,
\,\,\,\,\,\,\,\,\,\,\,\,\,\,\,\,\,\,\,\,\,\,\,
\cos\alpha\,=\,{mg_s\over [\mu_{(l,m)}]^{1/2}}\,
H_{(l,m)}^{-1/4}\,h_{(l,m)}^{1/2}\,\,.
\label{ciocho}
\eeq
The supersymmetry projection (\ref{cisiete}) certainly corresponds to that of a 
(F, D3) bound state of the type described above, with $\alpha$ being the mixing
angle.

\setcounter{equation}{0}
\section{Discussion}
\medskip
In order to check the stability of our configurations one can study their
behaviour under small fluctuations. This analysis, which we will not detail 
here, is similar to the one performed in refs. \cite{Bachas}-\cite{flux}
and shows that our
configurations are indeed stable. On the other hand, following ref.
\cite{flux}, one can verify that our solutions saturate a BPS bound on the
energy, which shows that they certainly minimize the energy. 

As compared to the cases studied in ref. \cite{flux}, it seems that the general
rule to find flux-stabilized configurations in a non-threshold bound state
background is to consider probes which are also extended in the directions
parallel to  the bound state in such a way that the probe could capture the flux
of the background gauge fields. However, nothing guarantees that the
corresponding configurations are  free of pathologies. To illustrate this fact,
let us consider the case of the background generated by  a (F,Dp) bound state. 
The string frame metric and dilaton for this bound state are \cite{FDp}:
\bear
ds^2&=&f_p^{-1/2}\,h_p^{-1/2}
\Big[\,-(\,dx^0\,)^2\,+\,(\,dx^{1}\,)^2\,+\,
\,h_p\,\Big((\,dx^{2}\,)^2\,+\,\cdots\,+\,\,(\,dx^{p}\,)^2\Big)\,\Big]\,+\rc\rc
&&+\,f_p^{1/2}\, h_p^{-1/2}\,
\Big[\,dr^2\,+\,r^2\,d\Omega_{8-p}^2\,\Big]\,\,,\rc\rc
e^{\tilde\phi}&=&f_p^{{3-p\over 4}}\,\,h_p^{{p-5\over 4}}\,\,,
\label{cinueve}
\eear
while the $B$ field is 
$B=\sin\varphi\,\,f_p^{-1}\,\,dx^{0}\wedge dx^{1}$ and the RR potentials are:
\bear
C^{(7-p)}_{\theta^1,\cdots,\theta^{7-p}}
&=&-\,\cos\varphi\,R^{7-p}\,C_p(\theta)\,
\sqrt{\hat g^{(7-p)}}\,\,, \rc\rc
C^{(9-p)}_{x^{0},x^1,\theta^1,\cdots, \theta^{7-p}}
&=&-\,\sin\varphi\,\cos\varphi\,R^{7-p}\,
f_p^{-1}\,C_p(\theta)\,
\sqrt{\hat g^{(7-p)}}\,\,.
\eear
According to our rule we should place a D(10-p)-brane 
probe extended along
$(t,x^{1},x^{2},r,\\\theta^1,\cdots,\theta^{7-p})$.
Moreover, we will adopt the ansatz
${\cal F}\,=\,{\cal F}_{0,1}\,dt\wedge dx^1\,
+\,  F_{2,r}\,dx^{2}\wedge dx^{r}$ for the gauge field, 
with ${\cal  F}_{0,1}\,=\,  F_{0,1}\,-\,f_p^{-1}\sin\varphi$ with constant
values of  $F_{0,1}$ and $F_{2,r}$. Following the same steps as in our previous
examples, we obtain that there exist  constant $\theta$ configurations if
$F_{0,1}\,=\,-\cos^2\varphi/\sin\varphi$. This is an overcritical field 
which makes negative the argument of the square root of the Born-Infeld term of
the action and, as a consequence, the corresponding value of $F_{2,r}$
is imaginary, namely  $F_{2,r}=-i\cos\theta$. These configurations are clearly
unacceptable. 
\medskip
\section{ Acknowledgments}
\medskip
We are grateful to A. Paredes and J. M. Sanchez de Santos for discussions. 
This work was
supported in part by DGICYT under grant PB96-0960,  by CICYT under
grant  AEN99-0589-CO2-02 and by Xunta de Galicia under  grant
PGIDT00-PXI-20609.


\begin{thebibliography}{99}

\bibitem{Bachas} C. Bachas, M. Douglas and C. Schweigert, 
{\sl \jhep} {\bf 0005}, 006 (2000), {\rm hep-th/0003037}.

\bibitem{Pavel} J. Pawelczyk,  {\sl \jhep} {\bf 0008}, 006 (2000), 
{\rm hep-th/0003057}.


\bibitem{PR} J. Pawelczyk and S.-J. Rey,  {\sl \pl} {\bf B493}
(2000) 395,  {\rm hep-th/0007154}.



\bibitem{flux}J. M. Camino, A. Paredes and A. V. Ramallo,
{\sl \jhep} {\bf 05} (2001) 011,  {\rm hep-th/0104082}.

\bibitem{Myers} R. C. Myers,  {\sl \jhep} {\bf 9912} (1999) 022, 
{\rm hep-th/9910053}; N. R. Constable, R. C. Myers and O. Tafjord, 
{\sl \pr} {\bf D61} (2000) 106009, {\rm hep-th/9911136}. 



\bibitem{NCback}J. C. Breckenridge, G. Michaud and R. C. Myers, 
{\sl \pr} {\bf D55} (1997) 6438, {\rm hep-th/9611174};
M. S. Costa and G. Papadopoulos, 
{\sl \np} {\bf B510} (1998) 217, {\rm hep-th/9612204}.


\bibitem{swedes}
M. Cederwall, A. von Gussich, B.E.W. Nilsson, P. Sundell and A.
Westerberg,
{\sl  Nucl. Phys.} {\bf B490} (1997) 179, {\rm hep-th/9611159};
E. Bergshoeff and P.K. Townsend, {\sl Nucl. Phys.} {\bf B490}
(1997) 145, {\rm hep-th/9611173};
M. Aganagic, C. Popescu and J.H. Schwarz, {\sl Phys. Lett.} {\bf B393} (1997)
311, {\rm hep-th/9610249}; {\sl
 Nucl. Phys.} {\bf B495} (1997) 99, {\rm hep-th/9612080}.




\bibitem{bbs}
K. Becker, M. Becker and A. Strominger, {\sl 
Nucl. Phys.} {\bf B456} (1995) 130, 
{\rm hep-th/9507158}; E. Bergshoeff, R. Kallosh, T. Ortin and G. Papadopulos,
{\sl  Nucl. Phys. } {\bf B502} (1997) 149, {\rm hep-th/9705040};
E. Bergshoeff and P.K. Townsend,  {\sl \jhep} {\bf 05} (1999) 021, 
{\rm hep-th/9904020}.

\bibitem{NS5Dp} J. X. Lu and S. Roy, {\sl  \pl} {\bf B248} (1998) 298, 
{\rm hep-th/9802080}; 
M. Alishahiha and Y. Oz, {\sl  \pl} {\bf B495} (2000) 418,  
{\rm hep-th/0008172};
M. Alishahiha, Y. Oz and J. Russo, {\sl \jhep} {\bf 09} (2000) 002,
{\rm hep-th/0007215};
I. Mitra and S. Roy, {\sl \jhep} {\bf 02} (2001) 026,
{\rm hep-th/0011236}. 


\bibitem{FDp}
J. Russo and A. A. Tseytlin, {\sl  \np} {\bf B490} (1997) 121, 
{\rm hep-th/9611047};
M. Green, N. D. Lambert, G. Papadopulos and P.K. Townsend,
{\sl  \pl} {\bf B384} (1996) 86,
{\rm hep-th/9605146};
J. X. Lu and S. Roy, {\sl  \np} {\bf B560} (1999) 181, 
{\rm hep-th/9904129}.






\end{thebibliography}
\end{document}